\documentclass[preprint,prl,showpacs,floatfix]{revtex4}

\usepackage{graphicx}
\usepackage{amsmath}

\begin{document}

\title
{
Excluded volume effect enhances the homology pairing of model chromosomes
}

\author
{
Kazunori Takamiya$^1$, Keisuke Yamamoto$^1$, Shuhei Isami$^1$, Hiraku Nishimori$^{1,2}$,  Akinori Awazu$^{1,2}$
}

\affiliation
{
$^1$Department of Mathematical and Life Sciences, Hiroshima University,
$^2$Research Center for Mathematics on Chromatin Live Dynamics. \\
Kagami-yama 1-3-1, Higashi-Hiroshima 739-8526, Japan.
}


\begin{abstract}
To investigate the structural dynamics of the homology pairing of polymers, we modeled the scenario of homologous chromosome pairings during meiosis in {\it Schizosaccharomyces pombe}, one of the simplest model organisms of eukaryotes. We consider a simple model consisting of pairs of homologous polymers with the same structures that are confined in a cylindrical container, which represents the local parts of chromosomes contained in an elongated nucleus of {\it S. pombe}. Brownian dynamics simulations of this model showed that the excluded volume effects among non-homological chromosomes and the transitional dynamics of nuclear shape serve to enhance the pairing of homologous chromosomes.
\end{abstract}


\maketitle

\section{Introduction}

Eukaryotes exhibit genetic recombination, i.e., the exchange of base pairs between chromosomes at homologous loci during meiosis, which helps to sustain their genetic diversity. This process requires synapsis formation between homologous loci along the lengths of maternal and paternal chromosomes. Recent theoretical studies suggest that the homologies of the sequence-dependent distributions of the electrostatic charge and the binding sites of DNA-bridging proteins play important roles in the recognition and pairing of homologous loci \cite{KL,REV_MOD,AF,KW,DL,VK,OOYAMA1,CT,KL1}.

Indeed, these recent studies can explain the mechanism of homology recognition among loci that  are already within a close distance to each other (i.e., $\sim nm$) during the processes of DNA damage repair and genetic recombination. On the other hand, in the early stage of meiosis, the initial distances between homologous loci on maternal and paternal chromosomes are often further than the nanometer scale, at distance closely matching the nuclear radius (i.e., $\sim \mu m$). Thus, to unveil the mechanism of the entire process of homology pairing for genetic recombination, the large-scale processes occurring in the entire nucleus, such as recognition of homologous chromosomes, should be considered before focusing on the above-mentioned nanometer scale processes. Therefore, in this study, we developed a simple model consisting of polymers inspired by the state of chromosomes during meiosis of the fission yeast {\it Schizosaccharomyces pombe} ({\it S. pombe}) to provide a possible mechanism underlying the pairing of homologous chromosomes.

{\it S. pombe} is one of the most popular model organisms of unicellular eukaryotes, and contains only three chromosomes \cite{pombe}. Although {\it S. pombe} cells are usually haploid, they often become diploid through zygote formation and exhibit genetic recombination during meiotic prophase, as illustrated in Fig. 1(a,b) \cite{meiosis1,hiraoka0,hiraoka1,hiraoka2}. Thus, this organism has been considered as an ideal model for experimental studies of chromosomal dynamics during meiosis.

In this paper, we consider a physical model of the polymers involved in the local parts of the chromosomes of {\it S. pombe} to unveil the mechanism of recognition between homologous chromosomes during meiosis. In the next section, a confined polymers system is introduced as a simple model of the chromosomes in the nucleus during meiosis based on recent experimental results. In the third section, we present the results of the developed model and consider the physical mechanism underlying the homology recognition of polymers and its generality. Finally, we provide an overall summary and the novel perspectives gained from this study.

\begin{figure}
\begin{center}
\includegraphics[width=7.0cm]{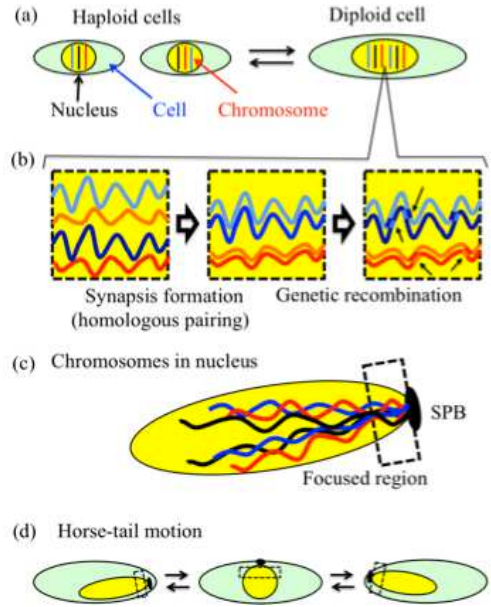}
\end{center}
\caption{(Color online) Illustrations of (a) haploid and diploid states, (b) homology pairing of chromosomes and genetic recombination during meiotic prophase, (c) elongated chromosomes in the elongated nucleus, and (d) "horse-tail motion" of the nucleus of {\it S. pombe}.}
\end{figure}

\section{Model}
\subsection{Assumptions of the model}
Based on recent experimental results, we constructed a model of chromosomes in the nucleus during the meiotic prophase of {\it S. pombe} according to the following two assumptions.

First assumption:

The distributions of the nucleosomes and the binding sites of DNA-associated proteins differ substantially among the 1st, 2nd, and 3rd chromosomes in {\it S. pombe} \cite{hiraoka2,nucp}, since they are highly dependent on DNA sequences that show vast variation among the three chromosomes. In general, the distributions of the nucleosomes and DNA-associated protein-binding sites determine the higher-order physical structure of chromatin \cite{hiraoka2,nucp,ooyama2}. This fact suggests that homologous chromosomes are structurally similar, whereas the structures of non-homologous chromosomes differ greatly. Furthermore, the fluctuations of such higher-order structures are negligibly smaller than the spatial scale of the local domains of a chromosome when several DNA-binding proteins function normally, such as a cohesive protein {e.g., \it Rec8} \cite{hiraoka2}. Thus, we assume that the local parts of a homologous chromosome have the same spatial structures, which are tightly maintained.

Second assumption:

During the entire period of meiotic prophase in {\it S. pombe}, the ends of chromosomes are clustered around the spindle pole body (SPB) on the nuclear membrane, and the SPB is continuously pulled by dynein on the cytoplasmic microtubules to help the nucleus move back and forth between the ends of the cell, resulting in the so-called "horse-tail motion" illustrated in Fig. 1 (c,d) \cite{meiosis1,hiraoka0,hiraoka1,hiraoka2}. This motion leads to elongation of the nucleus, and the relative force is exerted on each chromosome in the front-to-back direction. Thus, the chromosomes might also become elongated in the elongating nucleus. Accordingly, we assume that each local part of the chromosome is elongated by being pulled in the front-to-back direction, restricted by the area of the elongated space.

\begin{figure}
\begin{center}
\includegraphics[width=7.0cm]{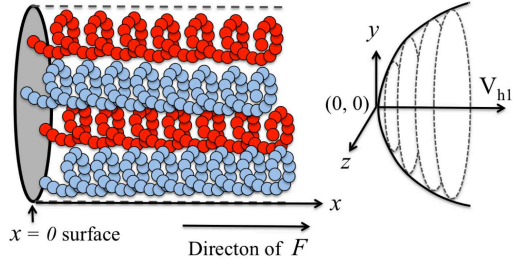}
\end{center}
\caption{(Color online) Schematic illustration of the model, showing the elongated polymer populations confined in an elongated three-dimensional container (left), and the potential $V_{hl}$ (right). The polymers shown in the same color represent homologous polymers with the same shape.}
\end{figure}

\subsection{Confined elongated polymers model}

Based on the assumptions listed above, a simple model of $M$ pairs of homologous elongated polymers confined in an elongated three-dimensional (3-D) container was developed, which mimics the local populations of chromosomes in the nucleus during the meiotic prophase of {\it S. pombe}. A schematic illustration of the present model is shown in Figure 2. We assume that the container and each polymer are elongated in the direction along the $x$ axis of the $x-y-z$ 3-D space, and that the center of the $y-z$ cross-section of the container is given by $(x, 0, 0)$. The $n$-th ($1 \le n \le 2M$) polymer is described by a chain consisting of $N^n$ particles with a unique basic structure. Here, as mentioned in Section 3.3, we consider one particle as a region containing $10 \sim 100$ kb nucleotides.

In the simulation model, each polymer is constructed using an elastic network model \cite{ela1,ela2,ela3}, in which some pairs of spherical particles are connected by springs based on their natural lengths, so that the basic structure of each polymer is stable. Here, the natural lengths of the springs between two centers of the neighboring particles are assumed to be equal to or slightly larger than the sums of their radii. Non-neighboring particles experience soft-core repulsion due to the excluded volume among them. We assume that the motion of one end particle of each polymer is restricted around $x=0$, which mimics the effect of the clustering of the ends of chromosomes around the SPB, and that the motions of all particles in the $y$ and $z$ directions are restricted by a potential originating from the restrictions of the nuclear membrane (see Fig. 2).

The equation of motion for each particle is given by
\begin{equation}
\gamma {\bf \dot{x}_i^n} = -\nabla_i (V_{int}(\{ {\bf x}_i^n, r_i^n \}) + V_{h}(\{ {\bf x}_i^n \}) ) + {\bf F}_p + {\bf \eta}_i^n(t),
\end{equation}
\begin{equation}
<{\bf \eta}_i^n(t){\bf \eta}_{i}^n(t')> = 2\gamma G\delta(t-t'),
\end{equation}
where ${\bf x}_i^n = (x_i^n, y_i^n, z_i^n)$ and $r_i^n$ are the position and radius, respectively, and ${\bf \eta}_i^n $ and $G$ are the random force and magnitude, respectively, working on the $i$-th particle in the $n$-th polymer. $\gamma$ indicates the coefficient of the drag force working on each particle.

The interaction potential among particles is given by $V_{int}(\{ {\bf x}_i^n \})=V^{ch}(\{ {\bf x}_i^n \}) + V^{sf}(\{ {\bf x}_i^n \})$. Here, the first term is the potential to stabilize the basic structure of each polymer as
\begin{equation}
V^{ch} =  \sum_{n}\sum_{i < j}\frac{k_{i,j}^n}{2}(|{\bf x}_i^n-{\bf x}_j^n|-L_{i,j}^n)^2,
\end{equation}
where $k_{i,j}^n$ and $L_{i,j}^n$ are the elastic constant and the distance between particles $i$ and $j$, respectively, of the basic structure of the $n$-th polymer. We set $k_{i,j}^n=k_{c_1}$ for $j = i+1$, $i+2$, and $i+3$; $k_{i,j}^n=k_{c_2}$ for $i+9$, $i+10$, and $i+11$; and $k^n_{i,j}=0$ otherwise. The second term indicates the effects of the excluded volumes of particles as
\begin{equation}
V^{sf} = \sum_{m \le n}\sum_{i < j}
\begin{cases}
\displaystyle
\frac{k_e}{2}(|{\bf x}_i^m-{\bf x}_j^n|-(r_i^m+r_j^n))^2 \,\,\,\,\,\,\,\,\, \left(|{\bf x}_i^m-{\bf x}_j^n| < r_i^m + r_j^n ,
\begin{cases}\,\,\,m < n \\
\,\,\,\,\,\,\,\,\,or \\
m = n, \, k_{i,j}^n=0
\end{cases}
\right)
\\
\\
  0 \,\,\,\,\,\, (Otherwise)
\end{cases}
\end{equation}
with elastic constant $k_e$. Here, we assume that the basic structure of the $n$-th polymer satisfies $r_i^n+r_j^n \le L_{i,j}^n$ for all $i$s and $j$s (see Appendices A and B).

$V_h(\{ {\bf x}_i^n \})$ indicates the potential of the container as
\begin{equation}
V_{h} = \sum_{n}\sum_{i} \frac{k_h(t)}{2}(|y_i^n|^2+|z_i^n|^2)+ \sum_{n} \frac{k_b}{2}|x_1^n|^2
\end{equation}
with restriction strength $k_h(t)$ induced by the nuclear membrane, and $k_b$ by the SPB for end particles. Here, we assume that the first term of Eq. (5) provides the influences of the nuclear membrane on chromosomes as explained below.

During the horse-tail motions, the nuclear membrane is soft and the nuclear width in the $y$ and $z$ directions tends to be reduced. Then, the chromosomes are affected by the force working in the direction of $y = z = 0$ by the membrane. Here, the influences of the membrane on the chromosome are expected to increase with an increase in the collision frequency between the chromosome and the membrane, which in turn seems to increase with an increase in the distance between the chromosome and the axis $y = z = 0$. Accordingly, in the following arguments, we assume that the average force working on each particle of chromosomes from the membrane is proportional to the distance between the particle position and the axis $y = z = 0$. In this case, the potential for the influences of the membrane on each particle is given by the first term of Eq. (5). It is noted that if we employ the potential with a more general form $\displaystyle \sum_{n}\sum_{i} \frac{k_h(t)}{a}\left(\sqrt{|y_i^n|^2+|z_i^n|^2}\right)^a$ instead of the first term of Eq. (5), qualitatively similar results are obtained, even if $a \ne 2$ for relatively small values of $a$.

${\bf F}_p = (F, 0, 0)$ indicates the force pulling the chromosomes in the front-to-rear direction, where $F$ given as $x^n_i > 0$ ($i>1$) always holds. In the following simulations, we use the parameter values $\gamma=1$, $G=300$ $k_e = 10000$, $k_{c_1} = 10000$, $k_{c_2} = 100$, $k_b = 1000$, and $F=20$. Here, we employ $k_e$ and $k_{c_1}$ as relatively large values compared to $G$, since the fluctuations of the higher-order structures of chromosomes are small, as mentioned in the first assumption listed in the previous section. It is noted that the results obtained in the next section are independent of the specific values of $k_e$, $k_{c_1}$, $k_{c_2}$, $k_b$, and $F$ if $k_e, k_{c_1} >> G$. The dependency of $\gamma$ and $G$ are considered in the last part of the next section.

\section{Results and Discussion}

\begin{figure}
\begin{center}
\includegraphics[width=6.0cm]{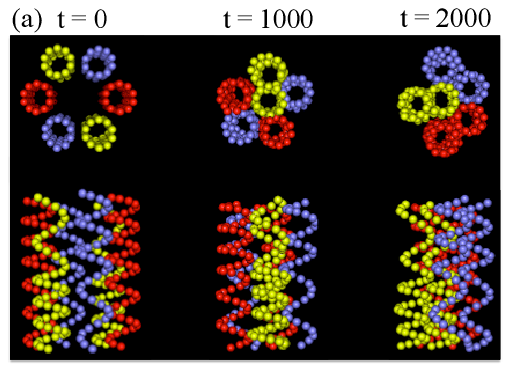}
\includegraphics[width=6.0cm]{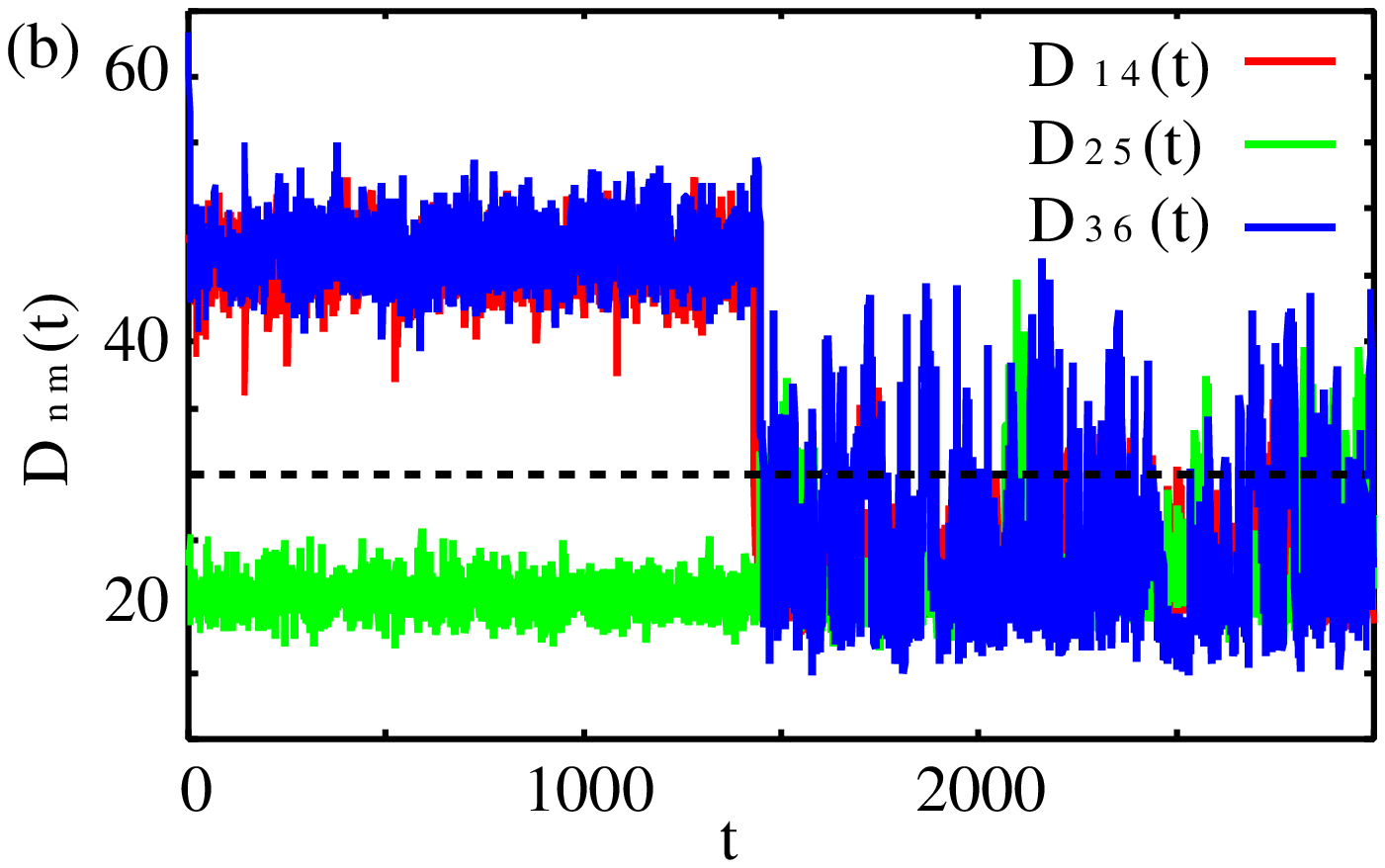}
\end{center}
\caption{(Color online) (a) Typical snapshots of the distribution of helical polymers (front views and top views) at the initial, intermediate, and final stages, and (b) corresponding $D_{nm}(t)$ between homologous polymers for the case of $k_{h}(t) = K = 1.5$. In (b), the pairing of green chromosomes of (a) is very fast, so that $D_{nm}(t)$ between them decreases and relaxes too quickly compared to the others.}
\end{figure}

\begin{figure}
\begin{center}
\includegraphics[width=6.0cm]{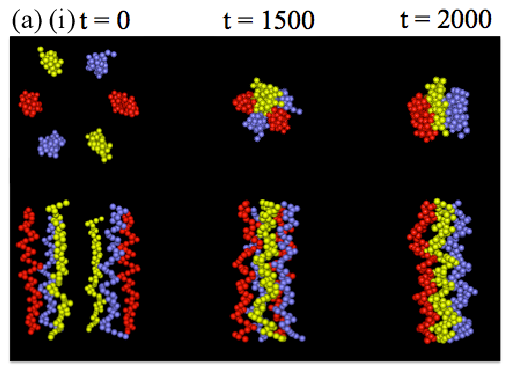}
\includegraphics[width=6.0cm]{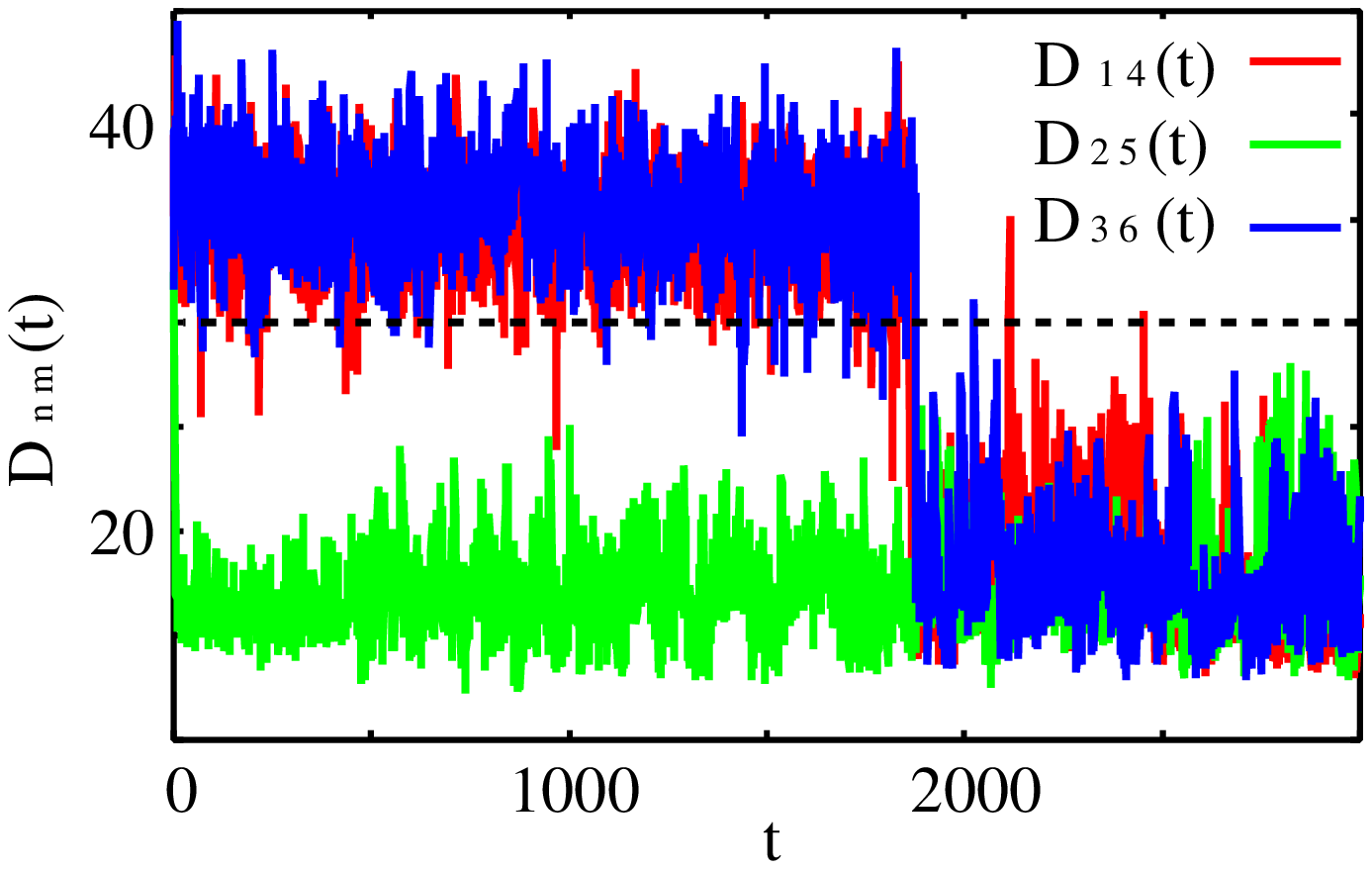}
\includegraphics[width=6.0cm]{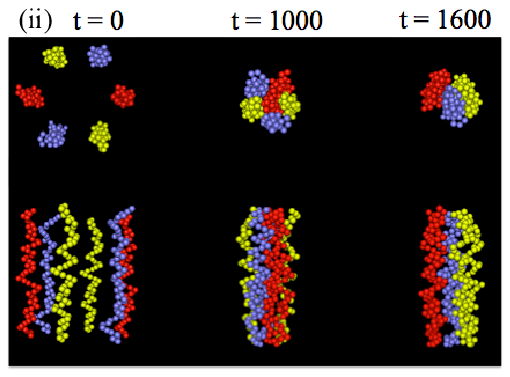}
\includegraphics[width=6.0cm]{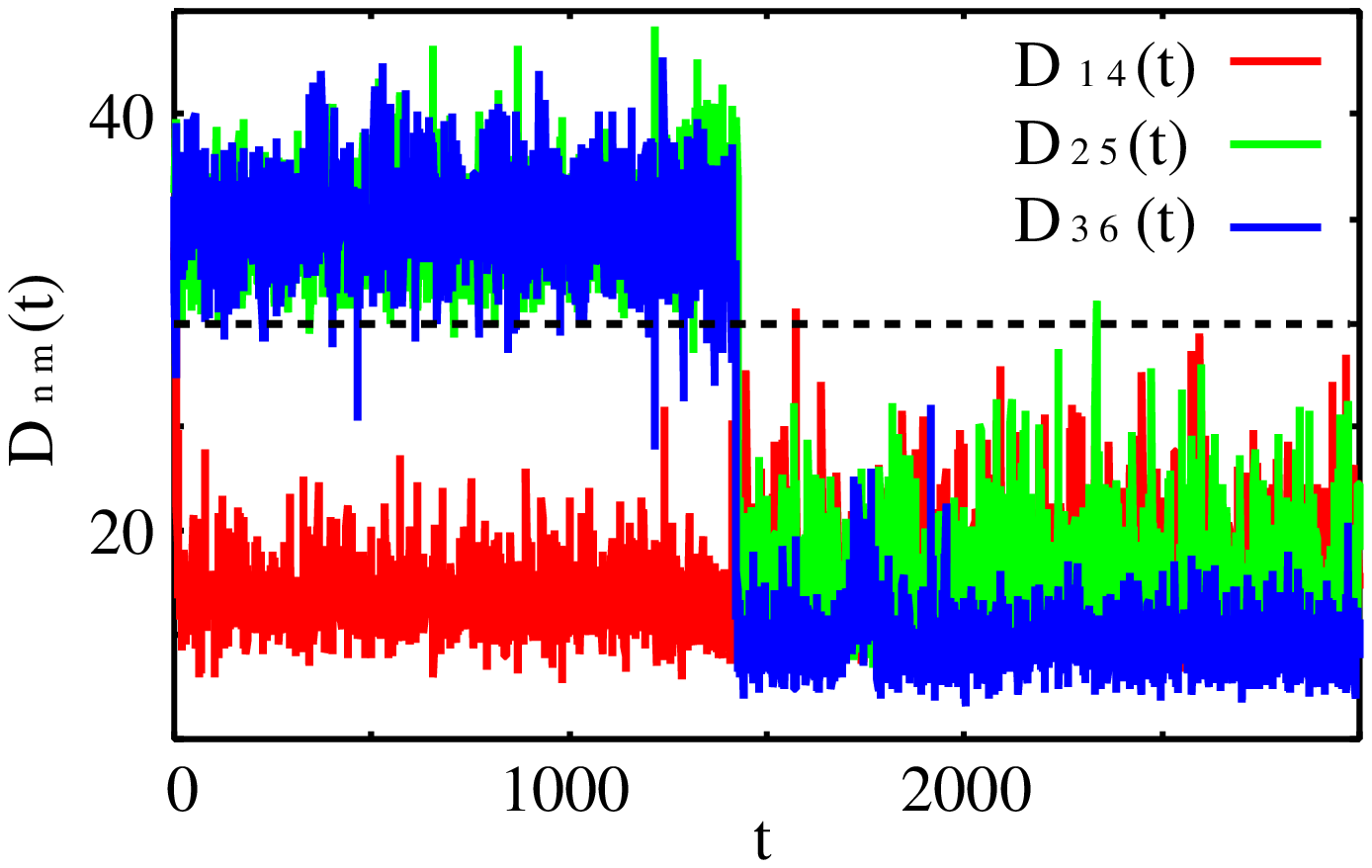}
\includegraphics[width=6.0cm]{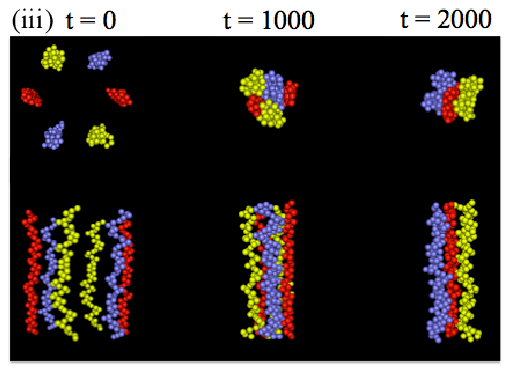}
\includegraphics[width=6.0cm]{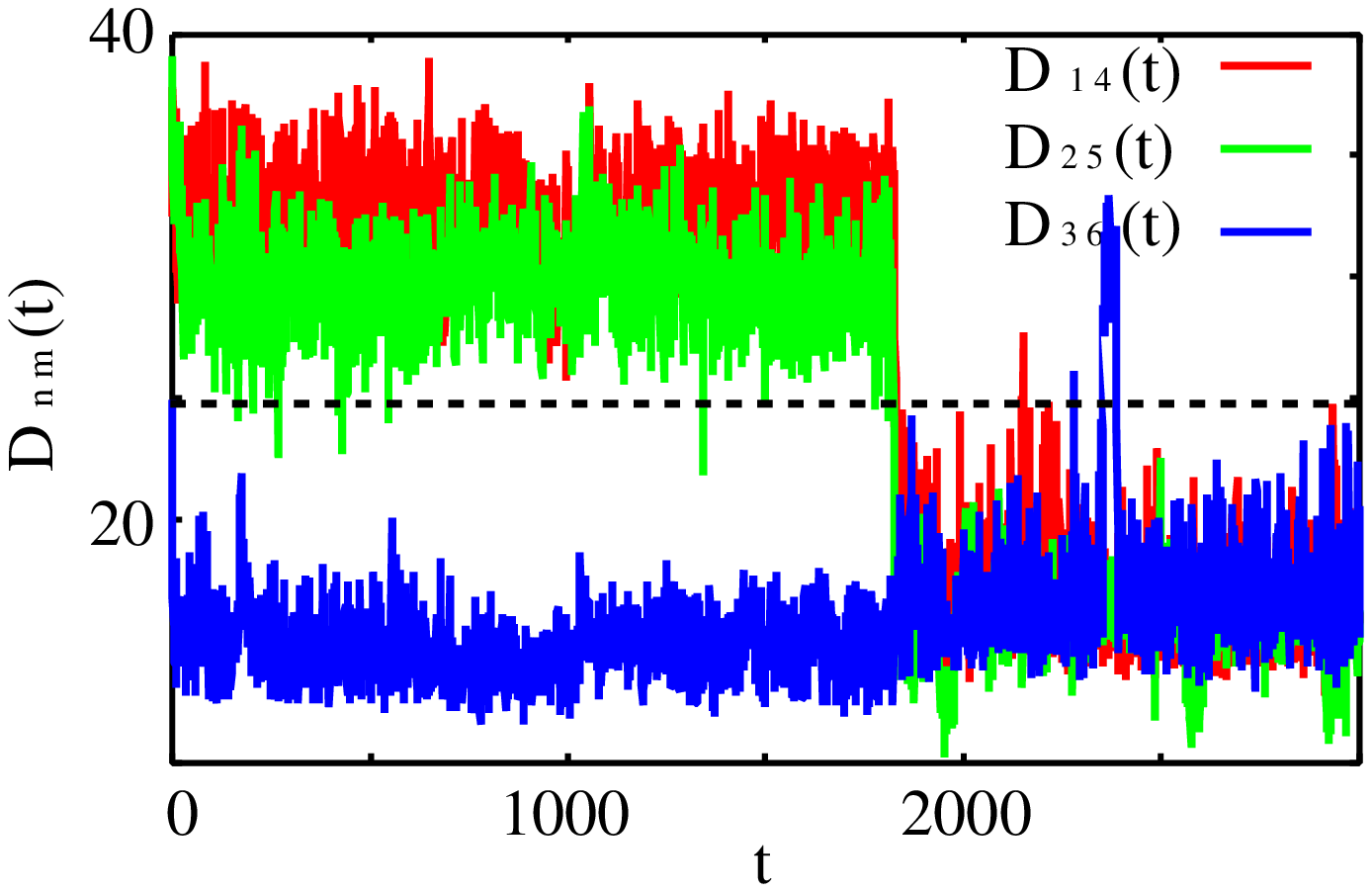}
\includegraphics[width=10.0cm]{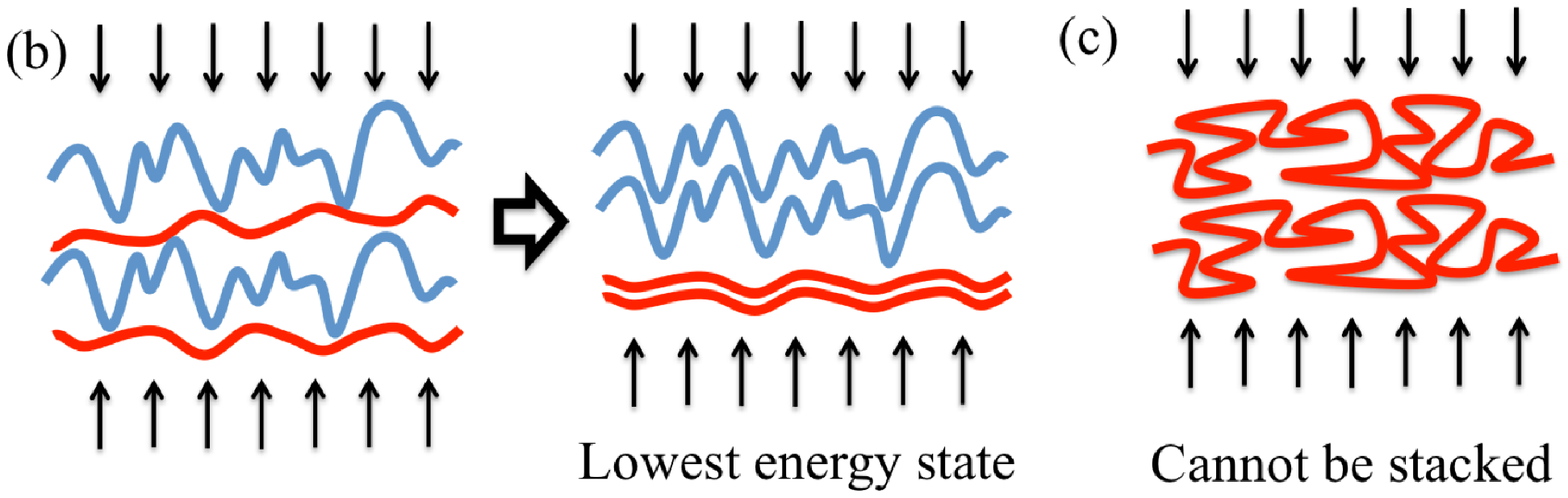}
\end{center}
\caption{(Color online) (a) Typical snapshots of the distribution of three homologous pairs of elongated random polymers (front views and top views) at the initial, intermediate, and final stages (Right), and corresponding $D_{nm}(t)$. (i), (ii), and (iii) indicate the results of three randomly selected random polymer populations with (i), (ii) $K=1.5$, and (iii) $K = 2.5$. The right panels in (a) indicate that the pairings of (i) green, (ii) red, (iii) blue chromosomes are very fast so that $D_{nm}(t)$ between them decreases and relaxes much more rapidly compared to the others. (b) Potential minimum state (Right) and non-minimum state (Left) of the system with pairs of elongated homologous polymers. (c) Polymers that are not sufficiently elongated cannot stack even if they are homologous.}
\end{figure}

\subsection{The excluded volume effect enhances homology pairing}
We here provide the results of the simulations of the present model, focusing on the spatial-temporal distributions of homologous and non-homologous polymers. Similar to the case of {\it S. pombe}, we consider a system consisting of three pairs of homologous polymers. Unfortunately, the specific structure of each chromosome during the meiotic prophase of {\it S. pombe} has not yet been elucidated in detail. Thus, as a first step, we employed one of the simplest 3-D structures as the basic polymer structure, where each polymer was constructed by the combination of helices with different wavelengths elongated in the positive $x$-axis direction. This model is referred to as the helical polymer model.

To construct the helical polymer model, we set the initial position of the center and the radius $r^n_i$ of each particle as described in Appendix A. Here, the initial helical structure of each polymer is regarded as its basic structure. We assume that the pairs of polymers $n=1,4$, $n=2,5$, and $n=3,6$ are homologous pairs, respectively, and set their initial positions to be relatively far apart.

First, we focus on the case in which $k_h(t)= K$ is constant. Figure 3(a) shows typical snapshots of the distribution of polymers at the initial, intermediate, and final stages in the case of $K = 1.5$. In this case, the homologous polymers become closer together over time, and in the final stage, the system relaxes to the state in which the homologous polymers are partially stacked on top of one another. In order to characterize the temporal evolution of the distance between each pair of homologous chromosomes, we measured $D_{nm}(t)=\sum_i^{N^n}|{\bf x}_i^n-{\bf x}_i^m|/N^n$, where the $n$-th and $m$-th polymers are the homologous polymers ($(n,m)=(1,4)$, $(2,5)$, and $(3,6)$, and $N^m = N^n$), ${\bf x}_i^n$, and ${\bf x}_i^m$ are the positions of the corresponding particles between these polymers. As shown in Fig. 3(b), $D_{nm}(t)$ for several pairs of homologous polymers tended to decrease with time, with some fluctuations. For the present helical polymers, the radius of each helix is $=10$ and the radius of each particle is given as $=3.1$ (See Appendix A). If the system does not contain any fluctuations, $D_{nm}(t)$ is expected to be smaller than $\sim 26.2$ when two homologous polymers are close together and stacked on top of each other. Thus, in the present simulations with finite fluctuations, we assume that two homologous polymers are close together when $D_{nm}(t)$ is smaller than $\sim 30$. Note that the qualitative properties of the system are independent of the details of this criterion. If we regard these polymers as the chromosomes during the meiotic prophase, this result suggests that several pairs of homologous loci can approach over time to eventually become close enough for recombination.

The mechanism contributing to this result can be easily understood given the model assumptions. In this model, we assumed that each polymer is constructed by the combination of elongated helices. Of note, two elongated helices with the same wavelength can be easily stacked, whereas two helices with different wavelengths cannot be stacked, even if the difference between their wavelengths is very small. Then, two central axes of two homologous polymers can be closer than those of non-homologous polymers, which means that the excluded volumes between homologous polymers are smaller than those among non-homologous polymers. Thus, the homology pairing of polymers can occur so as to construct a compact structure around the axis $y=z=0$ that minimizes the free energy of the system when the polymers are sufficiently restricted by the potential, indicating an influence of the nuclear membrane.

It should also be noted that the above-mentioned effects acting among homologous and non-homologous polymers are not limited to the helical polymer model but rather represent a general phenomenon for several polymer shapes, as long as they are sufficiently elongated. Figure 4(a) shows typical snapshots of the distributions of three homologous pairs of elongated random polymers at the initial, intermediate, and final stages, and $D_{nm}(t)$ for the three randomly selected sets of random polymer populations. The specific construction method of each random polymer is given in Appendix B. In general, each pair of homologous polymers tended to be stacked so as to construct a compact structure around the axis $y=z=0$ with selection of an appropriate $K$ value, since the free energy of the system reaches a minimum (Fig. 4(b,c)).

\subsection{Nuclear shape transitions also enhance homology pairing}

If $K$ is not set appropriately, the results described above cannot be obtained. Figure 5(a) shows the $D(t)= <(< D_{nm}(t) >_{nm})>_{samples}$ of the helical polymer model for $K$ values that are smaller or larger than the appropriate value ($K \sim 1.5$). Here, $<... >_{nm}$ indicates the average over all of the homologous polymers pairs ($(n,m)=(1,4)$, $(2,5)$, and $(3,6)$), and $<... >_{samples}$ indicates the average over 12 different simulation results using different random numbers generating $\eta^n_i(t)$. In the case of larger $K$ values, the pairing of homologous polymers still occurs but the process is slowed down and thus takes a much longer time than in the case with an appropriate $K$ value. In fact, with larger $K$ values, it is difficult to observe the pairing of homologous polymers since the simulation time is finite. On the other hand, for smaller $K$ values, $D_{nm}(t)$ does not converge, indicating that the stacked structure of each pair of homologous polymers is not stable, and thus the configuration of polymers changes frequently.

Moreover, recent experimental studies have shown that the nucleus goes through periodic phases of elongation and contraction due to the so-called ``horse-tail motion'' during the meiotic prophase of {\it S. pombe}, as shown in Fig. 1 \cite{meiosis1,hiraoka0,hiraoka1,hiraoka2}. Here, $\sim 30$ trips for the SPB between the front and rear ends of a cell were iterated, which induced $\sim 60$ elongations and contractions of the nucleus over two hours. When the nucleus was highly elongated, the spatial restriction from the nuclear membrane to the chromosomes became stronger in the vertical nuclear-traveling direction, but became weaker when the nucleus was contracted as shown in Fig. 1(d). Thus, it is naturally assumed that $k_h(t)$ varies periodically. Note that the period of such nuclear dynamics seems to be relatively slow compared to the diffusion time required for the local domains of chromosomes to pass through the free space with a length scale similar to their widths. Thus, to demonstrate the influences of the horse-tail motion on homology pairing, we performed simulations of the helical polymers model in which $k_h(t)$ oscillates slowly.

Figure 5 (b) and (c) show $D(t)$ values (12 samples) of the system consisting of three pairs of homologous helical polymers during $60$ iterations of $k_h(t)$ oscillations. Here, $k_h(t)=W(1+\sin 2\pi w t)$ is considered for several $W$ values with $w =0.01$. In this case, $D(t)$ exhibits large oscillations accompanying the oscillations of $k_h$. However, for $W$ values larger than $W \sim 2.5$, the average of $D(t)$ was smaller than $30$ after ten oscillations ($t > 1000$), where $D(t)<30$ holds for more than $\sim 70\%$ of the time, as shown in Fig. 5(d). Moreover, the decrease of the envelope of the lowest $D(t)$ for each oscillation is much faster than the decrease of $D(t)$ in the case that $k_h =K =$ constant, even if the averages of both $k_h$ values are the same. We obtained similar results over a wide range of $w$ values, except for much larger $w$.

It should be noted that homologous loci can often be bound by the electrostatic forces or the bindings of DNA-bridging proteins if these regions are sufficiently close, as mentioned in recent studies \cite{KL,REV_MOD,AF,KW,DL,VK,OOYAMA1,CT,KL1}. Thus, if homologous chromosomes are close enough for an appropriately long period of time, as observed in the present results for large $W$ values, the synapsis of homologous loci can form with a sufficiently high probability. These results suggest that the periodic structural changes of the nuclear membrane induced by the horse-tail motion of the nucleus can enhance the homology pairing of chromosomes.

\begin{figure}
\begin{center}
\includegraphics[width=6.0cm]{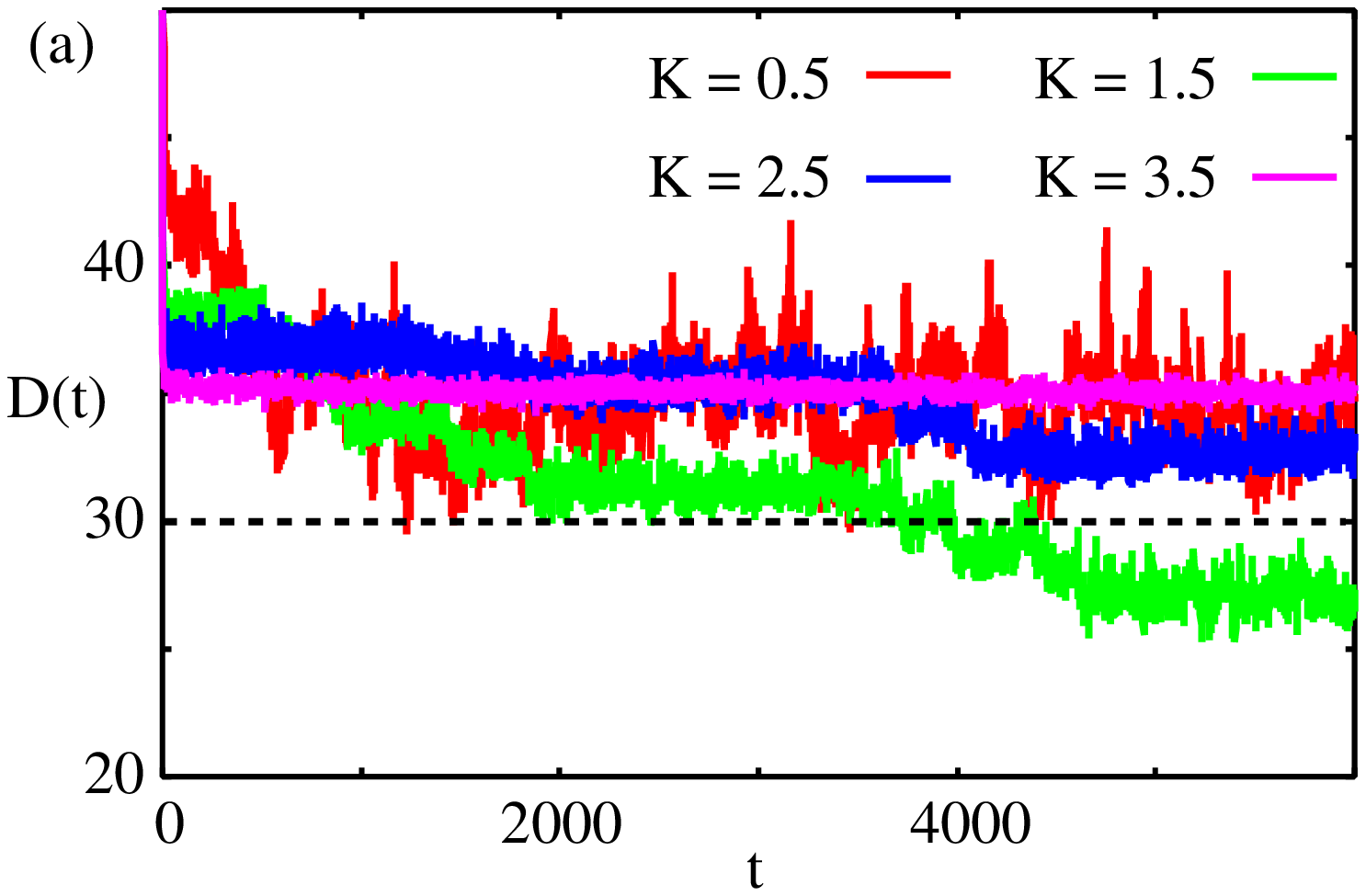}
\includegraphics[width=6.0cm]{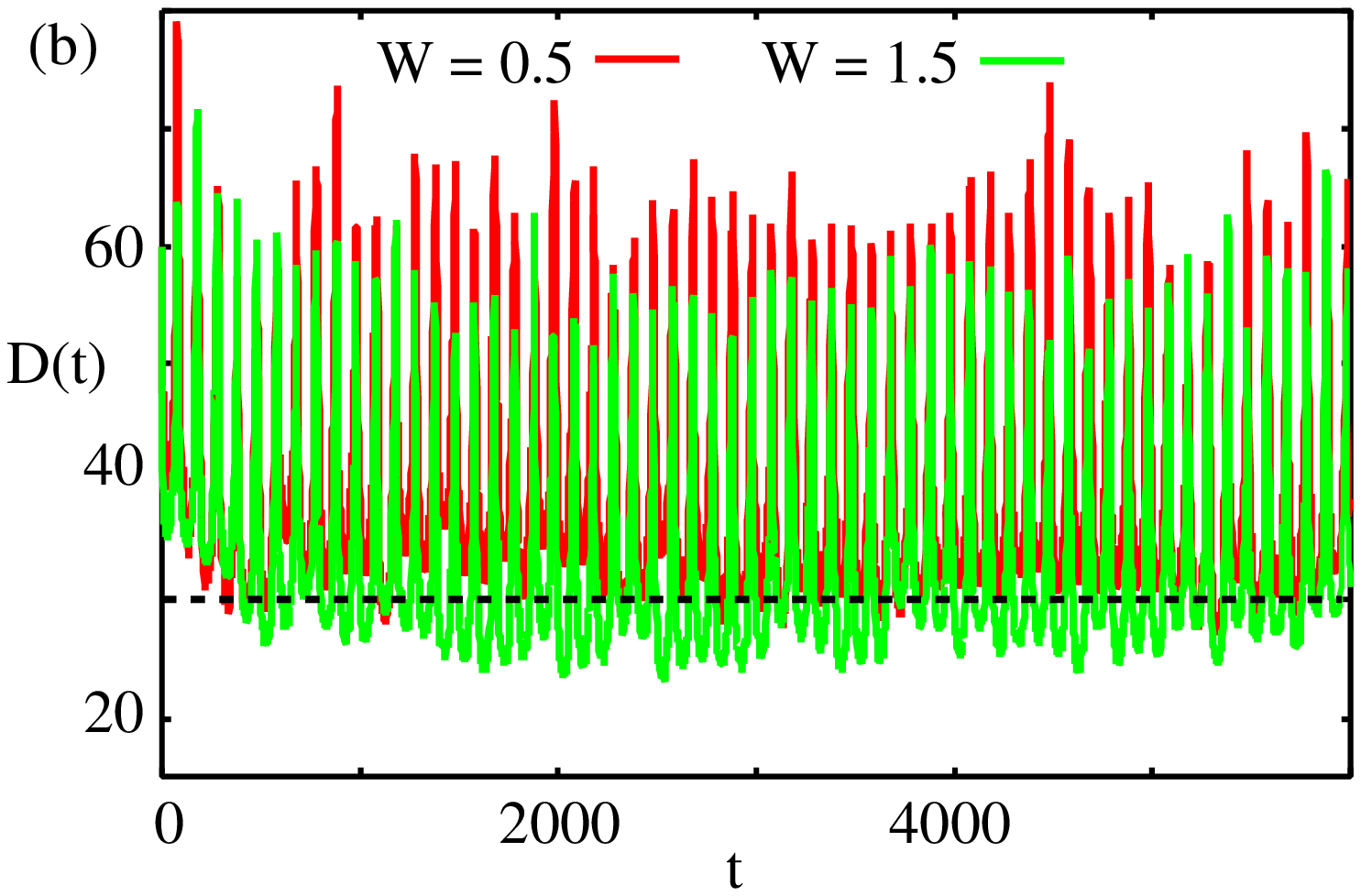}
\includegraphics[width=6.0cm]{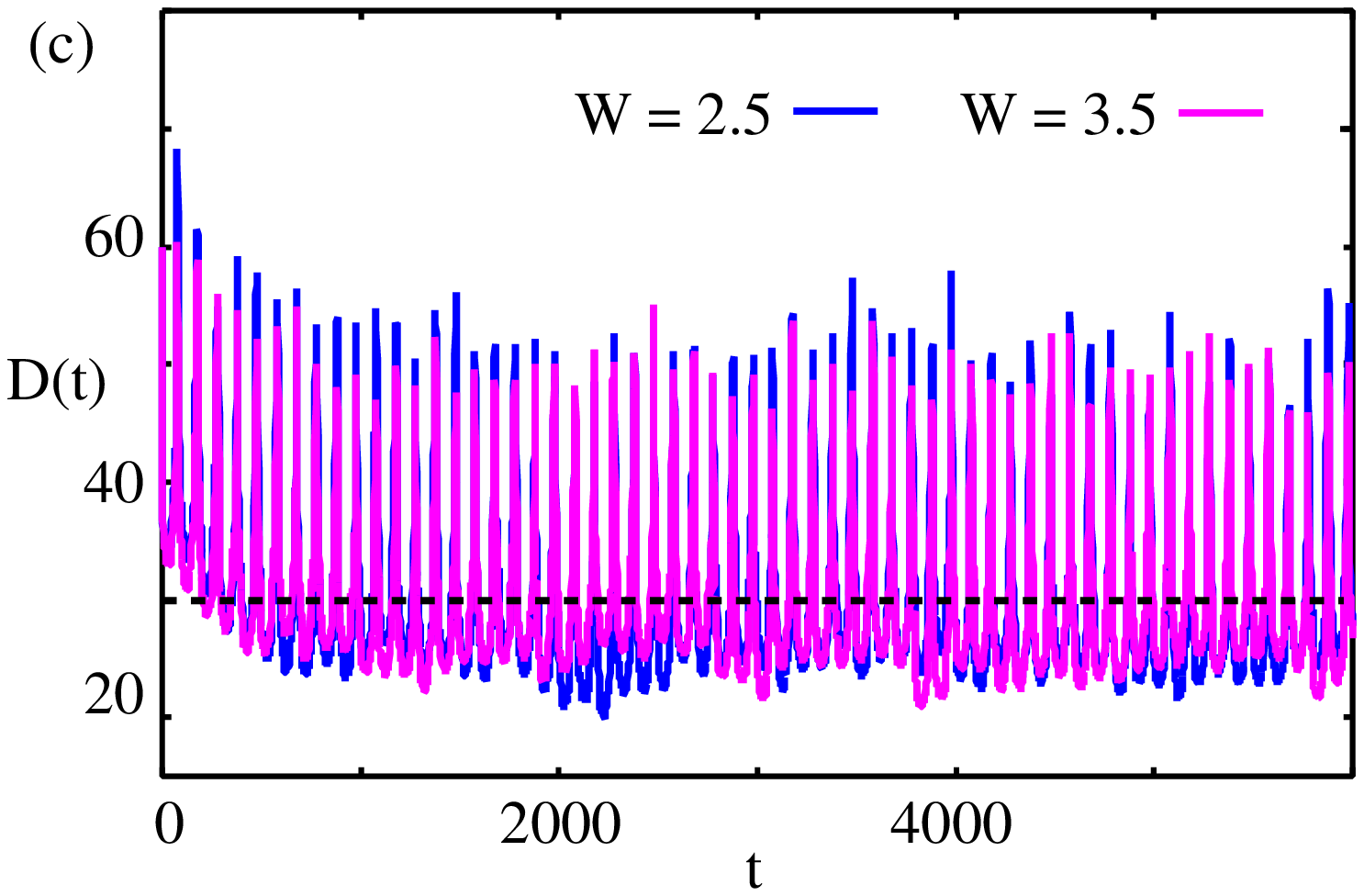}
\includegraphics[width=6.0cm]{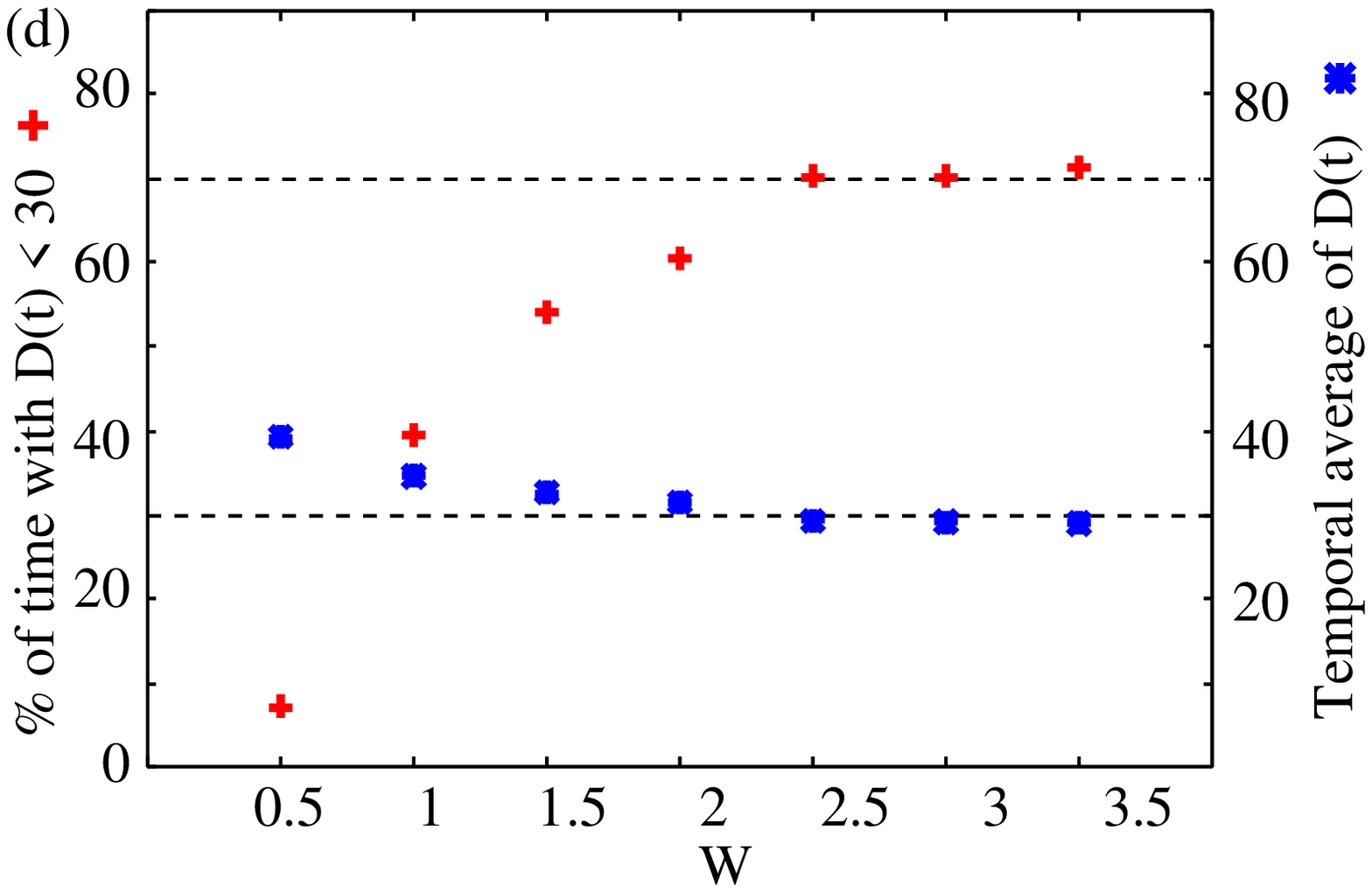}
\end{center}
\caption{(Color online) (a) $D(t)$ of the helical polymer model for the case of $k_{h}(t) = K = 0.5$, $1.5$, $2.5$, and $3.5$. (b, c) $D(t)$ of the helical polymer model for the case of $k_{h}(t) = W ( 1 + \sin 2\pi w t )$, with (b) $W=0.5$, $W=1.5$, and (c) $W=2.5$, considering $W=3.5$ with $w = 0.01$. (d) Percentage of time with $D(t)<30$ and temporal average of $D(t)$ for $t>1000$.}
\end{figure}

\subsection{Comparison between the model and experimental results}
In the present simulations, we employed simple values for several parameters in order to avoid technical complexity. Now, we compare the parameter values used for the simulations to those obtained from experimental results and observations in order to evaluate the applicability of the model. In particular, we focus on the ratio between the characteristic time of the temporal variations of the potential originating from the restrictions of the nuclear membrane $\tau_{n} \sim w^{-1}$ and that of the diffusion of each polymer in the restricted space $\tau_{d} \sim A^2/(G/\gamma)$, since the present simulations showed that the oscillations of $k_h$ and the diffusion of polymers are important factors for the homology pairing of chromosomes. In the present simulation, we assume $A \sim \sqrt{G/W}$, which indicates the scale of the radius of the space in which the particles can remain. Then, $\tau_{d} \sim 10^{-1 \sim 0}$, $\tau_n \sim 10^{1 \sim 2}$, and $\tau_n/\tau_d \sim 10^{1 \sim 3}$ are obtained.

Now, we assume that the radius of each particle in our model is similar to the radius of the complex of a ``30-nm fiber'' and DNA-binding proteins with an inertial radius of few tens of nanometers, where the 30-nm fiber represents the well-known characteristic intra-chromatin fiber structure with width $\sim 30 [nm]$ \cite{chro}. Then, the diameter of each particle $d$ may be assumed as $d \sim 10^{-7}[m]$ and involves $10 \sim 100$-kb nucleotides. Each polymer in our model contains $\sim 60$ particles, so that each polymer describes a DNA segment of $\sim$ megabase pairs that is similar in scale to a small part of the chromosomes in {\it S. pombe} ($12 \sim 35$ Mbp), as illustrated in Fig. 1.

The drag coefficient $\gamma$ for each particle with diameter $d \sim 10^{-7}[m]$ is estimated as $6 \pi \eta d/2 \sim 10^{-10}[kg \cdot s^{-1}]$, where $\eta \sim 10^{-4}[kg \cdot s^{-1} \cdot m^{-1}]$ is the viscosity of water. We assume that the order of $G$ is similar to $k_BT \sim 10^{-21}$ ($k_B$ is the Boltzmann constant), while the intra-nuclear environment should not be in equilibrium. Thus, $G/\gamma \sim 10^{-11} [m^2 \cdot s^{-1}]$. On the other hand, $A \sim 10^{-6}[m]$ and $\tau_n \sim 10^{2}[s]$ are expected, based on the experimental observations of the horse-tail motions of the nucleus. Thus, $\tau_d \sim 10^{-1}[s]$ is obtained, and $\tau_n/\tau_d \sim 10^{3}$ is expected.

The estimated $\tau_n/\tau_d$ for the experimental situations is similar to that obtained for the present simulations. Thus, we believe that the present simulation results can sufficiently describe similar behaviors to those of experimental situations in a qualitative manner.

\section{Summary and perspectives}
In this study, we developed a model of the dynamical features of local parts of chromosomes during meiosis of {\it S. pombe}. Based on the simulations of this model, we demonstrated the structural homology between each pair of homologous chromosomes and showed that the dynamical structural transition of the nucleus known as horse-tail motion plays an important role in the homology pairing of chromosomes.

Although we have mainly considered cases with simple polymers, our arguments can also be extended to more general cases involving populations of several elongated molecules. We are currently conducting experiments to obtain information on the detailed chromosome structures during the meiotic prophase of {\it S. pombe}, which will help to verify these arguments.

Moreover, the present arguments are not limited to the case of {\it S. pombe}, but are applicable to eukaryotes in general. For example, it is reasonable to expect that the first assumption of our model considering the relationship between the DNA sequence and the higher-order chromosome structure would be satisfied for several organisms. Indeed, the extremely huge oscillatory motions of the horse-tail motion of the nucleus are a specific phenomenon of {\it S. pombe}. However, recent studies in several eukaryotes have revealed several rotational and oscillatory motions of the nucleus during meiotic prophase, including in the rat, budding yeast {\it Saccharomyces cerevisiae}, and nematode {\it Caenorhabditis elegans} \cite{meiosis1,rat,yeast,celegance}. We expect that such active motions of the nucleus might play important roles in the pairing between homologous loci generally, and plan to extend our arguments to these organisms in the future.

The results of our previous study suggested that the interphase intra-nuclear chromosome positioning can also be affected by the nuclear active transitional motions \cite{awa}. Thus, the influences of the nuclear active motions on the organizations of intra-nuclear architectures seem to be important for several cells and cell stages. In the present arguments, we employed the simple potential to describe the influences of the nuclear membrane as a first step to consider the contributions of the nuclear active motions to the pairing of homologous chromosomes. However, the descriptions of the effects of nuclear motions need to be modified in more detail in order to study more realistic interactions between chromosomes and the nuclear membrane. We are currently conducting studies to address such issues by considering the mechanism underlying the organizations of more complex intra-nuclear architectures, and plan to report these results in the near future.
\\[3ex]

\noindent
{\bf Acknowledgments}

\noindent
The authors are grateful to T. Sugawara, S. Shinkai, M. Ueno, S. Tate, and Y. Hiraoka for fruitful discussions. This research is (partially) supported by the Platform Project for Supporting in Drug Discovery and Life Science Research (Platform for Dynamic Approaches to Living Systems) from the Ministry of Education, Culture, Sports, Science and Technology (MEXT) of Japan, the Japan Agency for Medical Research and Development (AMED), and the Grant-in-Aid for Scientific Research on Innovative Areas ``Spying minority in biological phenomena (No.3306) (24115515)'' and ``Initiative for High-Dimensional Data-Driven Science through Deepening of Sparse Modeling (No.4503)(26120525)'' of MEXT of Japan.
\\[3ex]

\noindent
{\bf Appendix A: Construction of the initial conditions for three pairs of helical polymers}

\noindent
In the case of a system consisting of three pairs of helical polymers, the initial configuration of particles is given as follows: the initial position of the $i$-th particle center $(i= 1 \sim N_f^n)$ in the $n$-th polymer is given as $(x_i^n, y_i^n, z_i^n) = (s_f^n (i-1), Y_o^n + A \cos(\omega i), Z_o^n + A \sin(\omega i))$, and that of the $j$-th particle $(j= N_f^n+1 \sim N^n)$ is given as $(x_j^n, y_j^n, z_j^n) = (s_f^n (N_f^n-1) + s_r^n (j-1-N_r^n), Y_o^n + A \cos(\omega j), Z_o^n + A \sin(\omega j))$ $(N^n = N_f^n+N_r^n)$, as shown in Fig. 3(a). Note that $s^n_f = s^{n'}_f$ and $s^n_r = s^{n'}_r$ hold for homologous polymers. We consider that polymers with $n=1,4$, $n=2,5$, and $n=3,6$ are homologous, and set their initial positions as $(Y_o^n, Z_o^n) = (B\cos(2\pi n/6),B\sin(2\pi n/6))$ so that they are initially far apart. In the present simulations, we use the parameter values $A = 10$, $\omega = \pi/5$, $s_f^1 = 2$, $s_f^2 = 13/9$, $s_f^3 = 13/4$, $s_r^1 = 13/9$, $s_r^2 = 13/4$, $s_r^3 = 2$, $N_f^1 = 26$, $N_f^2 = 36$, $N_f^3 = 16$, $N_r^1 = 36$, $N_r^2 = 16$, $N_r^3 = 26$, and $B=30$.

In the present simulations, we assume that the radius of the $i$-th particle in the $n$-th polymer is given by $r^n_i = r$, with $r = 3.1$. In this case, $r_i^n+r_j^n \le L_{i,j}^n$ always holds, by which the finite gap between each pair of neighboring particles tends to appear in the simulations. However, with the parameter values of $r^n_i$, $L_{i,j}^n$, and $k_{c1}$ given in the present simulations, such gaps are always narrow enough to hold $|{\bf x}_i^n-{\bf x}_{i+1}^n| - 2r << 2r$ for any $i$. In this case, no part of the polymer can pass through such gaps due to the excluded volume effect of each particle.
\\[3ex]

\noindent
{\bf Appendix B: Construction of the initial conditions for three pairs of random polymers}

\noindent
In the case of a system consisting of three pairs of random polymers, the structure of each polymer is constructed as follows. First, we set the $i$-th particle center in the $n$-th polymer ($N^n = N = const$) as $(x^n_i, y^n_i, z^n_i) = (2 r s i, Y_o^n+rand^y_i, Z_o^n + rand^z_i)$, where the radius of each particle is given by $r_i = r$, and $rand^y_i$ and $rand^z_i$ are random numbers with $rand^y_i \in [-5r, 5r]$ and $rand^z_i \in [-5r, 5r]$, respectively. Second, we solve the following equations
$$
{\dot y}_i^n = -\nabla_{y_i^n} V^{ch}
$$
$$
{\dot z}_i^n = -\nabla_{z_i^n} V^{ch}
$$
until the motion of all particles relaxes. The relaxed structure is considered as the basic structure of each polymer. We consider the polymers with $n=1,4$, $n=2,5$, and $n=3,6$ to be homologous and set the shapes of each homologous pair to be the same. The initial position of the $n$-th polymer is set as $(Y_o^n, Z_o^n) = (B\cos(2\pi n/6),B\sin(2\pi n/6))$ so that the homologous polymers are initially far apart. In the present simulations, we use the parameter values $r = 3.1$, $s = 1/3$, $N = 60$, and $B = 30$.

\end{document}